\documentclass[12pt,a4paper]{article}

\usepackage[latin1]{inputenc}
\usepackage{lineno}  
\usepackage{graphicx}  
\usepackage{xspace} 
\usepackage{booktabs} 
\usepackage{multirow}
\usepackage{color}
\usepackage{colortbl}
\usepackage{amsmath} 
\usepackage{ifthen} 
\usepackage{subfigure}
\usepackage{lscape} 

\usepackage{indentfirst}

\newboolean{pdflatex}
\setboolean{pdflatex}{true} 

\newboolean{articletitles}
\setboolean{articletitles}{true} 

\newboolean{uprightparticles}
\setboolean{uprightparticles}{false} 

\usepackage{amssymb}
\usepackage{amsfonts}
\usepackage{upgreek} 

\usepackage{ifpdf}

\usepackage{hyperref}    
\usepackage[all]{hypcap} 

\def\ux85 {UX85\xspace}

\ifthenelse{\boolean{uprightparticles}}%
{

 \def\Ppi         {\ensuremath{\uppi}\xspace}

 \def\PDelta      {\ensuremath{\Delta}\xspace}                 
 \def\PXi      {\ensuremath{\Xi}\xspace}                 
 \def\PLambda      {\ensuremath{\Lambda}\xspace}                 
 \def\PSigma      {\ensuremath{\Sigma}\xspace}                 
 \def\POmega      {\ensuremath{\Omega}\xspace}                 
 \def\PUpsilon      {\ensuremath{\Upsilon}\xspace}                 
 
 \def\PB      {\ensuremath{\mathrm{B}}\xspace}                 
                  
 \def\PD      {\ensuremath{\mathrm{D}}\xspace}

 \def\PK      {\ensuremath{\mathrm{K}}\xspace}

 \def\Pb      {\ensuremath{\mathrm{b}}\xspace}                 
 \def\Pc      {\ensuremath{\mathrm{c}}\xspace}

 \def\Pi      {\ensuremath{\mathrm{i}}\xspace}

 \def\Pu      {\ensuremath{\mathrm{u}}\xspace}

}
{

 \def\Ppi         {\ensuremath{\pi}\xspace}

 \mathchardef\PDelta="7101
 \mathchardef\PXi="7104
 \mathchardef\PLambda="7103
 \mathchardef\PSigma="7106
 \mathchardef\POmega="710A
 \mathchardef\PUpsilon="7107
                  
 \def\PB      {\ensuremath{B}\xspace}                 
                  
 \def\PD      {\ensuremath{D}\xspace}

 \def\PK      {\ensuremath{K}\xspace}

 \def\Pb      {\ensuremath{b}\xspace}                 
 \def\Pc      {\ensuremath{c}\xspace}

 \def\Pi      {\ensuremath{i}\xspace}

 \def\Pu      {\ensuremath{u}\xspace}

}

\def\uquark    {\ensuremath{\Pu}\xspace}

\def\cquark    {\ensuremath{\Pc}\xspace}

\def\bquark    {\ensuremath{\Pb}\xspace}

\def\pion  {\ensuremath{\Ppi}\xspace}

\def\pipm  {\ensuremath{\pion^\pm}\xspace}

\def\kaon  {\ensuremath{\PK}\xspace}
  \def\Kbar  {\kern 0.2em\overline{\kern -0.2em \PK}{}\xspace}

\def\Kz    {\ensuremath{\kaon^0}\xspace}
\def\Kzb   {\ensuremath{\Kbar^0}\xspace}
\def\KzKzb {\ensuremath{\Kz \kern -0.16em \Kzb}\xspace}
\def\Kp    {\ensuremath{\kaon^+}\xspace}
\def\Km    {\ensuremath{\kaon^-}\xspace}
\def\Kpm   {\ensuremath{\kaon^\pm}\xspace}

\def\KpKm  {\ensuremath{\Kp \kern -0.16em \Km}\xspace}

\def\Kstarz  {\ensuremath{\kaon^{*0}}\xspace}

  \def\Dbar    {\kern 0.2em\overline{\kern -0.2em \PD}{}\xspace}
\def\D       {\ensuremath{\PD}\xspace}

\def\Dz      {\ensuremath{\D^0}\xspace}
\def\Dzb     {\ensuremath{\Dbar^0}\xspace}
\def\DzDzb   {\ensuremath{\Dz {\kern -0.16em \Dzb}}\xspace}
\def\Dp      {\ensuremath{\D^+}\xspace}
\def\Dm      {\ensuremath{\D^-}\xspace}

\def\DpDm    {\ensuremath{\Dp {\kern -0.16em \Dm}}\xspace}

\def\Dstarz  {\ensuremath{\D^{*0}}\xspace}

\def\B       {\ensuremath{\PB}\xspace}
  \def\Bbar    {\kern 0.18em\overline{\kern -0.18em \PB}{}\xspace}

\def\Bpm     {\ensuremath{\B^\pm}\xspace}

\def\Bd      {\ensuremath{\B^0}\xspace}

  \def\Y#1S{\ensuremath{\PUpsilon{(#1S)}}\xspace}

\newcommand{\decay}[2]{\ensuremath{#1\!\to #2}\xspace}         

\def\to                 {\ensuremath{\rightarrow}\xspace}

\def\CP                {\ensuremath{C\!P}\xspace}

\def\AT#1     {\ensuremath{A_{\mathrm{T}}^{#1}}\xspace}           

\def\C#1      {\ensuremath{\mathcal{C}_{#1}}\xspace}                       
\def\Cp#1     {\ensuremath{\mathcal{C}_{#1}^{'}}\xspace}                    
\def\Ceff#1   {\ensuremath{\mathcal{C}_{#1}^{\mathrm{(eff)}}}\xspace}        
\def\Cpeff#1  {\ensuremath{\mathcal{C}_{#1}^{'\mathrm{(eff)}}}\xspace}       
\def\Ope#1    {\ensuremath{\mathcal{O}_{#1}}\xspace}                       
\def\Opep#1   {\ensuremath{\mathcal{O}_{#1}^{'}}\xspace}                    

\newcommand{\tev}{\ensuremath{\mathrm{\,Te\kern -0.1em V}}\xspace}
\newcommand{\gev}{\ensuremath{\mathrm{\,Ge\kern -0.1em V}}\xspace}
\newcommand{\mev}{\ensuremath{\mathrm{\,Me\kern -0.1em V}}\xspace}
\newcommand{\kev}{\ensuremath{\mathrm{\,ke\kern -0.1em V}}\xspace}
\newcommand{\ev}{\ensuremath{\mathrm{\,e\kern -0.1em V}}\xspace}
\newcommand{\gevc}{\ensuremath{{\mathrm{\,Ge\kern -0.1em V\!/}c}}\xspace}
\newcommand{\mevc}{\ensuremath{{\mathrm{\,Me\kern -0.1em V\!/}c}}\xspace}
\newcommand{\gevcc}{\ensuremath{{\mathrm{\,Ge\kern -0.1em V\!/}c^2}}\xspace}
\newcommand{\gevgevcccc}{\ensuremath{{\mathrm{\,Ge\kern -0.1em V^2\!/}c^4}}\xspace}
\newcommand{\mevcc}{\ensuremath{{\mathrm{\,Me\kern -0.1em V\!/}c^2}}\xspace}

\def\gsim{{~\raise.15em\hbox{$>$}\kern-.85em
          \lower.35em\hbox{$\sim$}~}\xspace}
\def\lsim{{~\raise.15em\hbox{$<$}\kern-.85em
          \lower.35em\hbox{$\sim$}~}\xspace}

\def\tell1  {TELL1\xspace}
\def\ukl1   {UKL1\xspace}

\newcolumntype{e}{@{ $\pm$ }l}         
\newcolumntype{R}{>{$}r<{$}}           
\newcolumntype{L}{>{$}l<{$}}           
\newcolumntype{C}{>{$}c<{$}}           
\newcolumntype{E}{@{ $\pm$ }>{$}l<{$}} 

\usepackage{wasysym}     

\usepackage{mciteplus}

\newcommand{\ifb}     {\ensuremath{\rm \, ifb}}

\newcommand{\re}[2][()] {\ifthenelse{\equal{#1}{()}}{{\ensuremath{{\rm \, Re}}\left(#2\right)}}
                                                    {{\ensuremath{{\rm \, Re}}\left[#2\right]}}}
\newcommand{\im}[2][()] {\ifthenelse{\equal{#1}{()}}{{\ensuremath{{\rm \, Im}}\left(#2\right)}}
                                                    {{\ensuremath{{\rm \, Im}}\left[#2\right]}}}

\newcommand{\afb}{A_{\bar{D}}}
\newcommand{\af} {\vphantom{\afb}A_D}

\newcommand{\CKM}{$C\!K\!M$\xspace}
\newcommand{\inbars}[1]{\left|#1\right|}
\newcommand{\inbarssq}[1]{\left|#1\right|^2}

\newcommand{\dpmcpket}{\left|     {D}_\pm^{\mathrm{CP}}      \right\rangle}

\newcommand{\kspipi}    {$K_s \pi^+ \pi^-$\xspace}
\newcommand{\kskk}      {$K_s K^+ K^-$\xspace}

\newcommand{\dzket}   {\left|       \vphantom{\bar{D}^0}     {D}^0      \right\rangle}
\newcommand{\dzbket}  {\left|                            \bar{D}^0      \right\rangle}

\newcommand{\mrow}[2]   {\multirow{#1}{*}{#2}}

\newcommand{\mcolr}[2]  {\multicolumn{#1}{r}{#2}}

\newcommand{\otoprule}  {\midrule[\heavyrulewidth]}

\newcommand{\bdk}    {\decay{\Bpm}{\D\Kpm}}
\newcommand{\bdpi}   {\decay{\Bpm}{\D\pipm}}
\newcommand{\bzdkst} {\decay{\Bd} {\D\Kstarz}}

\newcommand{\bdstk}  {\decay{\Bpm}{\Dstarz                   \Kpm}}

\newcommand{\dk}    {\D\Kpm}
\newcommand{\dpi}   {\D\pipm}
\newcommand{\zdkst} {\D\Kstarz}

\newcommand{\dstk}  {\Dstarz                   \Kpm}

\newcommand\uncheckbox{\makebox[0pt][l]{$\square$}\raisebox{.15ex}{\hspace{0.1em}\hphantom{$\checkmark$}}}

\begin{document}

\begin{titlepage}

\vspace*{-1.5cm}

\hspace*{-0.5cm}

\vspace*{4.0cm}

{\bf\boldmath\huge
\begin{center}
A strategy for a simultaneous measurement of \CP violation parameters related to the \CKM angle $\gamma$
in multiple $B$ meson decay channels
\end{center}
}

\vspace*{2.0cm}

\begin{center}
Jordi Garra Ticó$^1$
\bigskip\\
{\it\footnotesize
$ ^1$Cavendish Laboratory, University of Cambridge, Cambridge, United Kingdom\\
}
\end{center}

\vspace{\fill}

\begin{abstract}
  Several methods exist \cite{glw1,glw2,ads3,ggsz,td} to measure \CP violation observables related to
  the \CKM angle $\gamma$. These observables are different for every $B$ meson decay channel.
  However, the information they contain on $\gamma$ is encoded in a similar way for all of them.
  This paper describes a strategy for a simultaneous measurement including several $B$ meson decay
  channels, while taking into account any possible correlations between them.
\end{abstract}

\vspace*{2.0cm}
\vspace{\fill}

\end{titlepage}

\thispagestyle{empty}  

\newpage
\setcounter{page}{2}
\mbox{~}

\cleardoublepage

\newpage

\section{Introduction}
\label{sc.intro}

The CKM angle $\gamma$ is currently the least constrained of the angles of the unitarity triangle.
It is defined as
\[
\gamma = \arg\left( - \frac{V_{ud}V^\star_{ub}}{V_{cd}V^\star_{cb}} \right).
\]
Because $V_{ub}$ is the largest contribution to this phase, $\gamma$ is measured in decays that involve
an interference between the transitions $\bquark \to \uquark$ and $\bquark \to \cquark$, such as
$\Bpm \to \D^{(*)} K^{(*)\pm}$, $\Bd \to \D \Kstarz$ or, eventually in the future,
$\Bpm \to \D^{(*)} \pipm$, where $\D$ represents some admixture of the $D$ meson flavor eigenstates
\Dz\ and \Dzb, and the notation $(*)$ indicates a regular expression representing either one $*$ or none.

Several methods exist \cite{glw1,glw2,ads3,ggsz,td} to measure different \CP violation observables
that constrain the \CKM angle $\gamma$.
This paper presents an approach to simultaneously measure a reduced set of observables in multiple $B$
decay channels, independently of which of these methods is used for any $B$ decay considered.

\section{Unified approach}
\label{sc.approach}

\newcommand{\dsket}[1]{\left| \vphantom{\bar{D}^0}     {D}_{#1} \right\rangle}

\newcommand{\defaf}  {\left\langle f \left| \mathcal{H} \vphantom{D^0} \right| \! \vphantom{\bar{D}^0}     {D}^0     \right\rangle}
\newcommand{\defafb} {\left\langle f \left| \mathcal{H} \vphantom{D^0} \right| \!                      \bar{D}^0     \right\rangle}
\newcommand{\defap}  {\left\langle f \left| \mathcal{H} \vphantom{D^0} \right| \! \vphantom{\bar{D}^0}     {D}_+     \right\rangle}
\newcommand{\defam}  {\left\langle f \left| \mathcal{H} \vphantom{D^0} \right| \! \vphantom{\bar{D}^0}     {D}_-     \right\rangle}
\newcommand{\defapm} {\left\langle f \left| \mathcal{H} \vphantom{D^0} \right| \! \vphantom{\bar{D}^0}     {D}_\pm   \right\rangle}
\newcommand{\defapmc}{\left\langle f \left| \mathcal{H} \vphantom{D^0} \right| \! \vphantom{\bar{D}^0}     {D}_\pm^c \right\rangle}

\newcommand{\zpm} {z_\pm}
\newcommand{\zpmc}{z_\pm^c}

\subsection{The admixture coefficients $\zpmc$} \label{sc.formalism}

Several $B$ meson decay channels produce admixtures of neutral $D$ mesons that involve $\gamma$.
The notation for the $D$ meson states is $\dzket$ and $\dzbket$ for the flavor eigenstates and
$\dsket{\pm}$ for the $D$ meson produced in a $B$ decay. In this paper, $\dsket{+}$ denotes the
$D$ meson produced in a $B^+$ or $B^0$ decay, and $\dsket{-}$ denotes the $D$ meson produced in
a $B^-$ or $\bar{B}^0$ decay. A superscript $c$ is used to denote the $B$ decay channel.
In general, one can write
\begin{equation}
  \begin{aligned}
    \left| D_-^c \right\rangle &\sim \left|     {D}^0 \right\rangle + z_-^c \left| \bar{D}^0 \right\rangle, \\
    \left| D_+^c \right\rangle &\sim \left| \bar{D}^0 \right\rangle + z_+^c \left|     {D}^0 \right\rangle,
  \end{aligned}
  \Rightarrow
  \left\{
  \begin{aligned}
    A_-^c &\sim \af  + z_-^c \afb, \\
    A_+^c &\sim \afb + z_+^c \af ,
  \end{aligned}
  \right.
\end{equation}
where
$\af = \defaf$, $\afb = \defafb$ and $A_\pm^c = \defapmc$. The complex coefficients $\zpmc$ are generally
specific for every $B$ decay channel $c$, and are typically expressed in either Cartesian or polar coordinates as
\begin{equation}
  z_\pm^c = x_\pm^c + i\,y_\pm^c = r_c\, e^{i \delta} \, e^{\pm i \gamma},
\end{equation}
where all the parameters with a subscript or superscript $c$ are specific for every $B$ decay channel.
The total number of parameters in Cartesian coordinates for $N$ different $B$ decay channels is $4\,N$.

It should be noted that, by defining
\begin{equation} \label{eq.zc}
  z_c = r_c \, e^{i \delta_c},
\end{equation}
we have
\begin{equation}
  z_\pm^c = z_c \, e^{\pm i \gamma}.
\end{equation}
This reveals a clear channel invariant,
\begin{equation}
  \frac{z_+^c}{z_-^c} = e^{2 i \gamma}
  \quad \Rightarrow \quad
  \gamma = \frac{1}{2} \arg\left(\frac{z_+^c}{z_-^c}\right).
\end{equation}
If the $\zpmc$ coefficients are both multiplied by any complex coefficient $\xi$, the result will contain the
exact same information on $\gamma$. 
In particular, it is always possible to express the $\zpmc$ coefficients for channel $c$ as
\begin{equation}
  z_\pm^c = \xi_c \, z_\pm^{\mathrm{DK}},
\end{equation}
where
\begin{equation} \label{eq.xic}
  \xi_c = \frac{z_c}{z_{\mathrm{DK}}}.
\end{equation}

Notice that, by definition, from expressions (\ref{eq.zc}) and (\ref{eq.xic}), the $\xi_c$ coefficients do not depend
on $\gamma$ so they are formally \textbf{nuisance parameters}.

In order to simplify the notation, this paper uses $z_\pm = z_\pm^{\mathrm{DK}}$, where \textit{DK} refers to the
$\Bpm \rightarrow \D \Kpm$ decay mode, so that $z_\pm^c = \xi_c \, z_\pm$.

Performing a simultaneous fit for the Cartesian parameters using these $\xi_c$ coefficients
reduces the number of independent parameters in the fit from $4\,N$ to only $2\,(N+1)$ ($4$~for the $DK$
channel, and then only $2$ for each of the rest of the channels). This is only one more parameter than a
simultaneous fit for the polar coordinates, but it has the advantage that, with a Cartesian fit, the real
and imaginary components of $\zpm$ and $\xi_c$ are expected to exhibit Gaussian behavior.

\subsection{The $\eta$ function}

It is useful to define the $\eta$ function as
\begin{equation}
  \eta\left( a, b, \kappa \right) = \left| a \right|^2 + \left| b \right|^2 + 2 \kappa \re{ a^\star b },
\end{equation}
where $a,b \in \mathbb{C}$ and $\kappa \in \mathbb{R}$.
The $\kappa$ coefficient is called the \textit{coherence factor} because it gives an idea of the fraction of
coherent sum that contributes to $\eta$,
\begin{equation}
  \eta\left( a, b, \kappa \right) = \kappa \left| a + b \right|^2 + ( 1 - \kappa ) \left( \left| a \right|^2 + \left| b \right|^2 \right).
\end{equation}

In this paper, when the coherence factor argument is omitted it should be assumed that it is implicit.
If one of the complex arguments is omitted, it should be assumed that it is $1$. So,
\begin{equation}
  \eta\left( a \right) = \eta\left( a, 1, \kappa \right).
\end{equation}

The $\eta$ function has the following properties:
\begin{alignat}{2}
  \eta(a, b, \kappa) &= \eta(b, a, \kappa),                                 & \hspace{10mm}\text{Symmetric} \\
  \eta(a, b, \kappa) &= \inbarssq{a} \eta\left(1,\frac{b}{a},\kappa\right), & \hspace{10mm}\text{Scaling}
\end{alignat}

\subsection{Signal amplitude}

The signal probability distribution function is proportional to the squared amplitude integrated over a
certain region of the $B$ phase space,
\begin{equation}
  p_\pm^c \sim \int d\mathcal{P}_B \left| A_\pm^c \right|^2.
\end{equation}

If $A_c$ is the decay amplitude corresponding to a $b \rightarrow c$ transition and $A_u \, e^{\pm i \gamma}$ is
the decay amplitude corresponding to a $b \rightarrow u$ transition, then
\begin{align}
  A_- &\sim A_c \af  + A_u e^{-i \gamma} \afb, \\
  A_+ &\sim A_c \afb + A_u e^{+i \gamma} \af .
\end{align}

It is important not to forget that $A_c$ and $A_u$ are different for each $B$ decay channel, but to avoid
cluttering the following notation with too many indices, I skip them for just a moment.

In the case of a 2-body $B$ decay, such as $B^\pm \to D K^\pm$, the $B$ decay amplitudes $A_c$ and $A_u$ are
constants, and one can write
\begin{align}
  A_- &\sim \af  + z_- \afb, \label{eq.am} \\
  A_+ &\sim \afb + z_+ \af , \label{eq.ap}
\end{align}
where
\begin{equation}
  z_\pm = \frac{A_u}{A_c} e^{\pm i \gamma}.
\end{equation}

This implies that, for 2-body decays,
\begin{align}
  p_- &\sim \eta( \af , z_-\, \afb ), \\
  p_+ &\sim \eta( \afb, z_+\, \af  ).
\end{align}

In the case of a multibody $B$ decay with $3$ or more particles in the final state, such as $B^0 \to D K \pi$,
the amplitude $\left| A_\pm^c \right|^2$ is usually integrated over some region in the $B$ decay phase space
(around the $K^{*0}$ resonance for $B^0 \to D K \pi$).

By squaring the modulus of expressions (\ref{eq.am}) and (\ref{eq.ap}) and by defining
\begin{align}
  N_{\alpha\beta} &= \int d\mathcal{P}_B A^\star_\alpha A_\beta, \\
  X_{\alpha\beta} &= \frac{N_{\alpha\beta}}{\sqrt{N_{\alpha\alpha} N_{\beta\beta}}},
\end{align}
one can write
\begin{align}
  p_- 
      &\sim        \inbarssq{\af}  + \frac{N_{uu}}{N_{cc}} \inbarssq{\afb} +
            2 \inbars{X_{cu}} \re{\sqrt{\frac{N_{uu}}{N_{cc}}}\,\frac{X_{cu}}{\inbars{X_{cu}}} e^{-i\gamma} \af^\star \afb }, \\
  p_+ 
      &\sim        \inbarssq{\afb} + \frac{N_{uu}}{N_{cc}} \inbarssq{\af}  +
            2 \inbars{X_{cu}} \re{\sqrt{\frac{N_{uu}}{N_{cc}}}\,\frac{X_{cu}}{\inbars{X_{cu}}} e^{+i\gamma} \afb^\star \af }.
\end{align}

Notice that, because of the Cauchy-Schwarz inequality, $\left| X_{\alpha\beta} \right| \leq 1$.
Also, one can define
\begin{align}
  \kappa      &= \inbars{X_{cu}}, \\
  r           &= \sqrt{\frac{N_{uu}}{N_{cc}}}, \\
  e^{i\delta} &= \frac{X_{cu}}{\inbars{X_{cu}}}, \\
  z           &= r\, e^{i\delta}, \\
  z_\pm       &= z\, e^{\pm i \gamma},
\end{align}
so,
adding back the channel index and using the formalism described in \S\ref{sc.formalism}, the signal amplitude
probability distribution can always be expressed as
\begin{align}
  p^c_- &\sim \eta( \af , \xi_c \, z_- \, \afb, \kappa_c ) =
  \inbarssq{\af}  + \inbarssq{\xi_c \, z_-} \inbarssq{\afb} + 2 \, \kappa_c \, \re{ \xi_c \, z_- \af ^\star \afb }, \label{eq.pcm} \\
  p^c_+ &\sim \eta( \afb, \xi_c \, z_+ \, \af , \kappa_c ) =
  \inbarssq{\afb} + \inbarssq{\xi_c \, z_+} \inbarssq{\af}  + 2 \, \kappa_c \, \re{ \xi_c \, z_+ \afb^\star \af  }. \label{eq.pcp}
\end{align}

It should be noted that these expressions describe the physics of the admixture that leads to the
different \CP observables used to measure $\gamma$, but they are not specific to any method.

\newpage

\newcommand{\witheta}{\boolean{true}}

\section{Specific equations for established methodologies}
\label{sc.eqs}

\subsection{GLW equations}

The GLW method uses, on one hand, states that are accessible from only one of the flavor eigenstates,
either $\dzket$ or $\dzbket$, such that
$A^{    {D}}_{    {f}} = \left\langle                        f_{    {D}} | \mathcal{H} |                       {D}^0 \right\rangle$,
$A^{\bar{D}}_{\bar{f}} = \left\langle                        f_{\bar{D}} | \mathcal{H} |                   \bar{D}^0 \right\rangle$, but
$                        \left\langle                        f_{\bar{D}} | \mathcal{H} | \vphantom{\bar{D}^0}  {D}^0 \right\rangle =
                         \left\langle \vphantom{f_{\bar{D}}} f_{    {D}} | \mathcal{H} |                   \bar{D}^0 \right\rangle = 0$,
and, on the other hand, states that are accessible from one of the $CP$ eigenstates $\dpmcpket$, such that
$A^{\mathrm{CP}}_\pm = \left\langle f_\pm | \mathcal{H} | D^\mathrm{CP}_\pm \right\rangle$,
but $\left\langle f_\mp | \mathcal{H} | D^\mathrm{CP}_\pm \right\rangle = 0$.

The observables of interest are
\begin{align}
  R_{\mathrm{CP}}^{\pm \, c}
  &=
  \frac{\Gamma\left( B^- \rightarrow D_\pm^\mathrm{CP} h^- \right) +
        \Gamma\left( B^+ \rightarrow D_\pm^\mathrm{CP} h^+ \right)}
       {\Gamma\left( B^- \rightarrow     {D}^0         h^- \right) +
        \Gamma\left( B^+ \rightarrow \bar{D}^0         h^+ \right)} 
  =
  \frac{1}{2} \left| \frac{A^{\mathrm{CP}}_\pm}{A^D_f} \right|^2
  \ifthenelse{\witheta}
  {
    \frac{\eta\left( \pm \xi_c \, z_- \right) + \eta\left( \pm \xi_c \, z_+ \right)}
         {2},
  }
  {
    \frac{\left| 1 \pm \xi_c \, z_- \right|^2 + \left| 1 \pm \xi_c \, z_+ \right|^2}
         {2},
  }
  \\
  A_{\mathrm{CP}}^{\pm \, c}
  &=
  \frac{\Gamma\left( B^- \rightarrow D_\pm^\mathrm{CP} h^- \right) -
        \Gamma\left( B^+ \rightarrow D_\pm^\mathrm{CP} h^+ \right)}
       {\Gamma\left( B^- \rightarrow D_\pm^\mathrm{CP} h^- \right) +
        \Gamma\left( B^+ \rightarrow D_\pm^\mathrm{CP} h^+ \right)} 
  =
  \ifthenelse{\witheta}
  {
    \frac{\eta\left( \pm \xi_c \, z_- \right) - \eta\left( \pm \xi_c \, z_+ \right)}
         {\eta\left( \pm \xi_c \, z_- \right) + \eta\left( \pm \xi_c \, z_+ \right)}.
  }
  {
    \frac{\left| 1 \pm \xi_c \, z_- \right|^2 - \left| 1 \pm \xi_c \, z_+ \right|^2}
         {\left| 1 \pm \xi_c \, z_- \right|^2 + \left| 1 \pm \xi_c \, z_+ \right|^2}.
  }
\end{align}

\subsection{ADS equations}

The ADS method uses states that are accessible from both flavor eigenstates.
With the convention $CP \dzket = \dzbket$ and assuming no direct $CP$ violation in the $D$ decay,
one can define the $\rho$ ratio of amplitudes as
\begin{equation}
  \rho
  =
  \frac{A^{\bar{D}}_{    {f}}}
       {A^{    {D}}_{    {f}}},
\end{equation}
and
\begin{alignat}{2}
  \Gamma^\pm_\mathrm{fav}
  &=
  \Gamma\left( B^\pm \rightarrow D_\mathrm{fav} \, h^\pm \right)
  =
  \ifthenelse{\witheta}
  {
    \left| A^\pm_{\tilde{D}} A^D_f \right|^2
    \eta\left( 1, \rho \, \xi_c \, z_\pm \right)
  }
  {
    \left| A^\pm_{\tilde{D}} A^D_f \right|^2
    \left| 1 + \rho \, \xi_c \, z_\pm \right|^2
  }
  ,
  \\
  \Gamma^\pm_\mathrm{sup}
  &=
  \Gamma\left( B^\pm \rightarrow D_\mathrm{sup} \, h^\pm \right)
  =
  \ifthenelse{\witheta}
  {
    \left| A^\pm_{\tilde{D}} A^D_f \right|^2
    \eta\left( \rho, \xi_c \, z_\pm \right)
  }
  {
    \left| A^\pm_{\tilde{D}} A^D_f \right|^2
    \left| \rho + \xi_c \, z_\pm \right|^2
  }
  ,
\end{alignat}
where $D_\mathrm{sup}$ and $D_\mathrm{fav}$ refer to suppresed and favored decay modes of
the produced $D$ meson.

The observables of interest are
\begin{equation}
  R_{\mathrm{ADS}}^{\pm \, c}
  =
  \frac{\Gamma^\pm_\mathrm{sup}}
       {\Gamma^\pm_\mathrm{fav}}
  =
  \ifthenelse{\witheta}
  {
    \frac{ \eta\left( \rho,         \xi_c \, z_\pm \right) }
         { \eta\left( 1   , \rho \, \xi_c \, z_\pm \right) }.
  }
  {
    \left|
      \frac{ \rho +         \xi_c \, z_\pm }
           { 1    + \rho \, \xi_c \, z_\pm }
    \right|^2.
  }
\end{equation}

It is also frequent to use the observables
\begin{alignat}{2}
  R_{\mathrm{ADS}}^c
  &=
  \frac{\Gamma^-_\mathrm{sup} + \Gamma^+_\mathrm{sup}}
       {\Gamma^-_\mathrm{fav} + \Gamma^+_\mathrm{fav}}
  &&=
  \ifthenelse{\witheta}
  {
    \frac{\eta\left( \rho, \xi_c \, z_- \right) +
          \eta\left( \rho, \xi_c \, z_+ \right)}
         {\eta\left(1, \rho \, \xi_c \, z_- \right) +
          \eta\left(1, \rho \, \xi_c \, z_+ \right)},
  }
  {
    \frac{\left| \rho +         \xi_c \, z_- \right|^2 +
          \left| \rho +         \xi_c \, z_+ \right|^2}
         {\left| 1    + \rho \, \xi_c \, z_- \right|^2 +
          \left| 1    + \rho \, \xi_c \, z_+ \right|^2},
  }
  \\
  A_{\mathrm{ADS}}^{\mathrm{sup} \, c}
  &=
  \frac{\Gamma^-_\mathrm{sup} - \Gamma^+_\mathrm{sup}}
       {\Gamma^-_\mathrm{sup} + \Gamma^+_\mathrm{sup}}
  &&=
  \ifthenelse{\witheta}
  {
    \frac{\eta\left( \rho, \xi_c \, z_- \right) -
          \eta\left( \rho, \xi_c \, z_+ \right)}
         {\eta\left( \rho, \xi_c \, z_- \right) +
          \eta\left( \rho, \xi_c \, z_+ \right)},
  }
  {
    \frac{\left| \rho + \xi_c \, z_- \right|^2 -
          \left| \rho + \xi_c \, z_+ \right|^2}
         {\left| \rho + \xi_c \, z_- \right|^2 +
          \left| \rho + \xi_c \, z_+ \right|^2},
  }
  \\
  A_{\mathrm{ADS}}^{\mathrm{fav} \, c}
  &=
  \frac{\Gamma^-_\mathrm{fav} - \Gamma^+_\mathrm{fav}}
       {\Gamma^-_\mathrm{fav} + \Gamma^+_\mathrm{fav}}
  &&=
  \ifthenelse{\witheta}
  {
    \frac{\eta\left( 1, \rho \, \xi_c \, z_- \right) -
          \eta\left( 1, \rho \, \xi_c \, z_+ \right)}
         {\eta\left( 1, \rho \, \xi_c \, z_- \right) +
          \eta\left( 1, \rho \, \xi_c \, z_+ \right)}.
  }
  {
    \frac{\left| 1 + \rho \, \xi_c \, z_- \right|^2 -
          \left| 1 + \rho \, \xi_c \, z_+ \right|^2}
         {\left| 1 + \rho \, \xi_c \, z_- \right|^2 +
          \left| 1 + \rho \, \xi_c \, z_+ \right|^2}.
  }
\end{alignat}

\subsection{GGSZ equations}

The GGSZ method uses $3$ or more body decays to final states that can be accessed from any of
$\dzket$ or $\dzbket$. As opposed to GLW or ADS, this method does not involve intermediate
observables, and the goal is to fit for the \CP observables, using equations
(\ref{eq.pcm}-\ref{eq.pcp}) directly.

\subsection{Time dependent equations}

The time evolution of the amplitude of the $B_s \to D_s K$ decay is governed by the equation
\begin{equation}
  A^{B_s}_{D_s^\pm}(t)
  = A^{B_s}_{D_s^\pm} \left[ g_+(t) + \lambda_\pm g_-(t) \right],
\end{equation}
where $t$ is the $B_s$ meson lifetime and $g_\pm(t)$ are the functions that describe the mixing of
the $B$ meson flavor eigenstates. The $\lambda_\pm$ parameters can be expressed as
\begin{equation}
  \lambda_\pm = \xi_c \, z_\pm \, e^{ \pm i \phi_q },
\end{equation}
where the channel $c$ is $B_s \to D_s K$ in this case.

\newcommand{\lambdafsq}{\left|\lambda_f\right|^2}
\newcommand{\lambdamsq}{\left|\lambda_-\right|^2}

Unfortunately, current measurements of $\gamma$ in decays with $B$ meson mixing introduce intermediate
parameters instead of targeting the $z^c_\pm$ observables themselves. In these cases, the squared
amplitude of the time-dependent probability distribution function is expressed as
\begin{equation}
  \small
  e^{-\Gamma t} \,
  \frac{1 + \lambdamsq}{2} 
  \left[\vphantom{\frac{1 - \lambdamsq }{1 + \lambdamsq}}\right.
    \cosh(x \Gamma t) +
    \underbrace{\frac{1 - \lambdamsq   }{1 + \lambdamsq}}_{\displaystyle C_f}                \cos (y \Gamma t)-
    \underbrace{\frac{-2 \re{\lambda_-}}{1 + \lambdamsq}}_{\displaystyle A_f^{\Delta\Gamma}} \sinh(x \Gamma t)-
    \underbrace{\frac{ 2 \im{\lambda_-}}{1 + \lambdamsq}}_{\displaystyle S_f}                \sin (y \Gamma t)
    \left.\vphantom{\frac{1 - \lambdamsq}{1 + \lambdamsq}}\right],
\end{equation}
and similarly for the conjugated amplitude, where $x$ and $y$ are parameters that describe the $B$ meson
mixing and $\Gamma$ is its decay rate. This approach makes this kind of analysis observable based, but
even in this case, it is possible to introduce the $C_f$, $A_f^{\Delta\Gamma}$ and $S_f$ parameters, and
the similar parameters for the conjugate decay, in a simultaneous fit that shares the $z_\pm$ parameters
with other $B$ decay channels and only introduces $\xi_c$ as a new coefficient.

\section{Experimental considerations}

There are several advantages in a simultaneous measurement, compared to a combination of standalone measurements:
\begin{itemize}
  \item It allows to properly take into account the correlations introduced by crossfeed components that are
    signal in one channel but background in another.
  \item For $N$ $B$ decay channels, the number of parameters in the fit is reduced from $4 N$ to $2 N + 2$ in
    Cartesian coordinates or $2 N + 1$ in polar coordinates. The minimum number of physical parameters is $2 N + 1$
    and, introducing just one additional redundant parameter, all the observables exhibit Gaussian behavior.
  \item Systematic uncertainties can also be evaluated simultaneously, with $B$ decay channel correlations
    properly taken into account.
\end{itemize}

\newpage

\newcommand{\bl}{\hphantom{-}}

\newcommand{\DKstz}{{D\hspace{-1pt}K\hspace{-1pt}^{\star\hspace{-0.5pt}0}}}
\newcommand{\DKst} {{D\hspace{-1pt}K\hspace{-1pt}^{\star}}}
\newcommand{\DstK} {{D\hspace{-0.7pt}^\star\hspace{-1.5pt} K}}

\section{Sensitivity study}

As a proof of concept, I have used \texttt{cfit} \cite{cfit} to generate $1000$ experiments of toy
Monte Carlo signal events for the $4$ $B$ decay channels $\bdk$, $\bzdkst$, $\bdstk$ and $\bdpi$, and
for the $2$ $D$ decay channels \kspipi and \kskk, using world averages from UTfit \cite{utfit}.
Since there is no information on the admixture coefficients for $\bdpi$, and considering that, for
this channel, they are expected to be one order of magnitude smaller than for $\bdk$, I used
$\xi_{D\pi} = 0.08 + 0.06 \, i$. The size of the toy experiments has been estimated for $3\,\ifb$ of
LHCb data, using world average branching fractions and reasonable guesses for the efficiencies.

\newcommand{\floatkappa}{\mcolr{1}{float $\kappa$}}
\newcommand{\fixkappa}  {\mcolr{1}{fix   $\kappa$}}

I have used the GGSZ method to perform a fit to the \CP observables $z_\pm$ and $\xi_c$ presented in \S\ref{sc.formalism}.
To ensure convergence, I have used a triple fit strategy, with the following fit steps:
\begin{enumerate}
  \item Fit for $z_\pm$ to the $DK$ sample only. \label{it.fix}
  \item Fix $z_\pm$ from the previous step 
    and fit for $\xi_{DK^{\star0}}$, $\xi_{D^\star K}$ and $\xi_{D\pi}$ to the whole sample.
  \item Float all parameters using the previous fit results as initial values.
\end{enumerate}
All the fits for the $z_\pm$ and $\xi_c$ parameters have been obtained using \texttt{cfit} \cite{cfit}.

\newcommand{\dxm}    {\delta x_-}
\newcommand{\dym}    {\delta y_-}
\newcommand{\dxp}    {\delta x_+}
\newcommand{\dyp}    {\delta y_+}
\newcommand{\drxDkst}{\re{\delta \xi_{DK^{\star0}}}}
\newcommand{\dixDkst}{\im{\delta \xi_{DK^{\star0}}}}
\newcommand{\drxDstk}{\re{\delta \xi_{D^{\star}K}}}
\newcommand{\dixDstk}{\im{\delta \xi_{D^{\star}K}}}
\newcommand{\drxDpi} {\re{\delta \xi_{D\pi}}}
\newcommand{\dixDpi} {\im{\delta \xi_{D\pi}}}
\newcommand{\sxm}    {\sigma_{x_-}}
\newcommand{\sym}    {\sigma_{y_-}}
\newcommand{\sxp}    {\sigma_{x_+}}
\newcommand{\syp}    {\sigma_{y_+}}
\newcommand{\srxDkst}{\re{\sigma_{\xi_{DK^{\star0}}}}}
\newcommand{\sixDkst}{\im{\sigma_{\xi_{DK^{\star0}}}}}
\newcommand{\srxDstk}{\re{\sigma_{\xi_{D^\star K}}}}
\newcommand{\sixDstk}{\im{\sigma_{\xi_{D^\star K}}}}
\newcommand{\srxDpi} {\re{\sigma_{\xi_{D\pi}}}}
\newcommand{\sixDpi} {\im{\sigma_{\xi_{D\pi}}}}
\newcommand{\blank}  {\mcolr{1}{\mbox{---}}}

\newcommand{\tr}[1]{\mrow{2}{$#1$}}
\newcommand{\trblank}{\mrow{2}{\mbox{---}}}

A standalone fit to $\bdk$ is used as reference, and several simultaneous fits are obtained by progressively adding more channels
to the approach. The obtained statistical uncertainties and biases are summarised in table \ref{tb.results}, where all collections
include $\bdk$ and the \kspipi and \kskk $D$ decay channels are considered.

\begin{table}[h]
\renewcommand{\tabcolsep}{1mm}
{\fontsize{7pt}{8.4pt}\selectfont
\begin{tabular}{lRRRRRRRRRR}
  \toprule
  collection & \sxm        & \sym        & \sxp        & \syp        & \srxDpi     & \sixDpi     & \srxDkst   & \sixDkst   & \srxDstk & \sixDstk \\
  \otoprule
  $\dk$      &     0.0172  &     0.0186  &     0.0210  &     0.0240  &     \blank  &     \blank  &     \blank &     \blank & \blank   & \blank   \\
  \midrule
  $\dpi$     &     0.0175  &     0.0185  &     0.0212  &     0.0245  &     0.0673  &     0.0794  &     \blank &     \blank & \blank   & \blank   \\
  \midrule
  $\zdkst$   &     0.0149  &     0.0169  &     0.0211  &     0.0236  &     \blank  &     \blank  &     2.11   &     2.26   & \blank   & \blank   \\
  \midrule
  $\dpi$     & \tr{0.0151} & \tr{0.0170} & \tr{0.0211} & \tr{0.0235} & \tr{0.0648} & \tr{0.0750} & \tr{2.09}  & \tr{2.26}  & \trblank & \trblank \\
  $\zdkst$   &             &             &             &             &             &             &            &            &          &          \\
  \midrule
  $\dstk$    &      0.0146 &     0.0169  &     0.0211  &     0.0239  &     \blank  &     \blank  &     \blank &     \blank & 0.442    & 0.426    \\
  \midrule
  all        &      0.0132 &     0.0159  &     0.0208  &     0.0233  &     0.0654  &     0.0739  &     2.02   &     2.28   & 0.429    & 0.418    \\
  \bottomrule
  \\
  \toprule
  collection & \dxm          & \dym          & \dxp          & \dyp          & \drxDpi     & \dixDpi      & \drxDkst    & \dixDkst    & \drxDstk & \dixDstk \\
  \otoprule
  $\dk$      &     -0.00198  &     -0.00209  &     -0.00225  &     -0.00178  &     \blank  &     \blank   &     \blank  &     \blank  & \blank   & \blank   \\
  \midrule
  $\dpi$     &     -0.00207  &     -0.00077  &     -0.00154  &     -0.00109  &     0.0133  &     0.00917  &     \blank  &     \blank  & \blank   & \blank   \\
  \midrule
  $\zdkst$   &     -0.00184  &     -0.00141  &     -0.00185  &     -0.00117  &     \blank  &     \blank   &     -0.149  &     -0.411  & \blank   & \blank   \\
  \midrule
  $\dpi$     & \tr{-0.00162} & \tr{-0.00059} & \tr{-0.00154} & \tr{-0.00096} & \tr{0.0137} & \tr{0.00782} & \tr{-0.150} & \tr{-0.448} & \trblank & \trblank \\
  $\zdkst$   &               &               &               &               &             &              &             &             &          &          \\
  \midrule
  $\dstk$    &     -0.00207  &      0.00018  &     -0.00138  &     -0.00072  &     \blank  &     \blank   &     \blank  &     \blank  & -0.00980 & -0.0206  \\
  \midrule
  all        &     -0.00165  &      0.00065  &     -0.00143  &     -0.00058  &     0.0144  &     0.00739  &     -0.126  &     -0.394  &  0.00078 & -0.0091  \\
  \bottomrule
\end{tabular}
}
\caption{Statistical uncertainties (upper part) and biases (lower part) of all the parameters involved in each fit.
  All collections include $\bdk$ plus the indicated $B$ decay channels.} \label{tb.results}
\end{table}

It should be noted that including $\bdpi$ does not improve the sensitivity on $\gamma$ with the used sample size.
However, each other additional channel contributes significantly to a smaller uncertainty on the $(x,y)_\pm$
parameters. All the biases obtained on any parameter are at least an order of magnitude smaller than the
statistical uncertainty on that parameter.

\newpage

\section{Conclusions}

I have presented an approach to measure the \CP parameters $z_\pm$ simultaneously in different $B$ decay channels.
This approach allows to properly take into account the correlations introduced by crossfeed components that are
signal in one channel but background in another, it significantly reduces the number of parameters in the fit and
allows a common treatment of the systematic uncertainties.

A more detailed study with realistic backgrounds expected at LHCb will follow.

\section*{Acknowledgements}

I wish to thank the Science and Technology Facilities Council (STFC, UK).
I am very thankful to my colleagues at Cambridge, Valerie Gibson, Susan Haines and
Chris Jones, for very interesting discussions.

\addcontentsline{toc}{section}{References}
\bibliographystyle{format/LHCb}
\bibliography{bib/biblio}

\end{document}